# The Relativistic Stern-Gerlach Force

C. Tschalär

## 1. Introduction

For over a decade, various formulations of the Stern-Gerlach (SG) force acting on a particle with spin moving at a relativistic velocity in an electromagnetic field have been put forward [1] and experiments proposed. To answer these speculations, the SG interaction, including the effect of the Thomas precession, have been derived from the well-established non-relativistic SG interaction energy using the Lagrangian formalism.

For a particle of mass $M$, charge $e$, spin $\vec{s}$, and gyro-magnetic factor $g$, the Lagrangian $L^*$ in the rest frame (RF) of the particle is [2]

$$L^* = -Mc^2 - e\phi^* + L^*_{SG}$$

where the asterisk indicates RF quantities, $\phi^*$ is the electro-magnetic scalar potential in the RF, and

$$L^*_{SG} = \vec{s} \cdot \vec{\Omega}^*$$

is the component of the RF Lagrangian (referred to in the following for short as "Lagrangian") representing the spin interaction with the electro-magnetic field [4]. Here, $\vec{\Omega}^*$ is the angular velocity of the spin precession around the magnetic field $\vec{B}^*$ in the RF described by [3]

$$d\vec{s} / dt^* = \vec{s} \times \vec{\Omega}^*$$

where

$$\vec{\Omega}^* = \frac{e}{M} \cdot \frac{g}{2} \vec{B}^* \quad .$$

## 2. Transformation to the Laboratory Frame

The RF is moving with velocity $c\vec{\beta}$ with respect to the laboratory frame (LF). The electro-mechanical interaction Lagrangian $L^*_{EM} = -Mc^2 - e\phi^*$ obeys the requirement that the product $\gamma L$ be invariant under Lorentz transformations [2] such that $\gamma L = \gamma^* L^* = L^*$ where $\gamma = 1/\sqrt{1-\beta^2}$ and $\gamma^* = 1/\sqrt{1-\beta^{*2}} = 1$, so that the electro-mechanical Lagrangian $L_{EM}$ in the LF is

$$L_{EM} = -Mc^2 / \gamma - e\phi + e\vec{\beta} \cdot \vec{A}$$

where $\phi$ and $\vec{A}$ are the scalar and vector potentials of the electro-magnetic field in the LF.
The transformation of the spin motion $d\vec{s} / dt^*$ in the RF to the LF has been derived in ref. [3] using the 4-vector representation $S^\alpha$ of the spin $\vec{s}$ :

$$S^\alpha = \left[ \gamma\left(\vec{\beta} \cdot \vec{s}\right);\ \vec{s} + \frac{\gamma^2}{\gamma+1}\left(\vec{\beta} \cdot \vec{s}\right)\vec{\beta} \ \right]$$

which, in the RF, becomes

$$S^{\alpha*} = \left[0; \vec{s}\right] \quad .$$

The result is [3] that the spin motion in the LF becomes

$$d\vec{s}/dt = \left(\vec{s}\times\vec{\Omega}^*\right)/\gamma - \vec{s}\times\vec{\omega}_T \equiv \vec{s}\times\vec{\Omega} \tag{1}$$

where $\vec{\omega}_T$ is the Thomas precession of the RF with respect to the LF if the particle, and with it, its RF, is accelerated perpendicularly to $\vec{\beta}$:

$$\vec{\omega}_T = -\frac{e}{M}\left(\frac{1}{\gamma+1}\right)\left(\vec{\beta}\times\vec{E}^*/c\right)$$

where $\vec{E}^*$ is the electric field in the RF. Therefore, the spin precession velocity $\vec{\Omega}$ in the LF is

$$\vec{\Omega} = \vec{\Omega}^*/\gamma - \vec{\omega}_T = \frac{e}{M}\left[\frac{g}{2\gamma}\vec{B}^* + \frac{1}{\gamma+1}\left(\vec{\beta}\times\vec{E}^*/c\right)\right] \quad .$$

The "Lagrangian" $L_{SGT}$ of the spin interaction with the electro-magnetic field in the LF is therefore

$$L_{SGT} = \vec{s}\cdot\vec{\Omega} = L^*_{SG}/\gamma - \vec{s}\cdot\vec{\omega}_T$$

which, expressed in electro-magnetic fields $\vec{E}$ and $\vec{B}$ in the LF, is

$$L_{SGT} = \frac{e}{M}\left\{\left(\frac{g}{2}-\frac{\gamma}{\gamma+1}\right)\vec{\beta}\left(\vec{s}\times\vec{E}/c\right) - \left(\frac{g}{2}-1\right)\frac{\gamma}{\gamma+1}\left(\vec{\beta}\vec{s}\right)\left(\vec{\beta}\vec{B}\right) + \left(\frac{g}{2}-1+\frac{1}{\gamma}\right)\left(\vec{s}\vec{B}\right)\right\}. \tag{2}$$

$L_{SGT}$ is the well-know spin interaction "Lagrangian" for a particle in an electro-magnetic field [4]. While the transformation of $L^*_{SG}$ to $L_{SGT}$ is thus modified by the Thomas precession of the RF, the electro-mechanical interaction Lagrangian $L_{EM}$ is not affected by this rotation of the RF.

## 3. Canonical Momentum

The canonical momentum $\vec{P}$ for the Lagrangian $L$ is defined as [5]

$$c\vec{P} \equiv \frac{\partial L}{\partial\vec{\beta}} = \gamma\vec{\beta}Mc^2 + e\vec{A} + \frac{\partial L_{SGT}}{\partial\vec{\beta}} = \gamma\vec{\beta}Mc^2 + e\vec{A} + c\vec{P}_{SGT} \quad . \tag{3}$$

The notation $\partial/\partial\vec{\beta}$ denotes the vector $(\partial/\partial\beta_1\ ;\ \partial/\partial\beta_2\ ;\ \partial/\partial\beta_3)$. The Stern-Gerlach-Thomas part $c\vec{P}_{SGT} = \partial L_{SGT}/\partial\vec{\beta}$ of $c\vec{P}$ is, according to eq. (2),

$$\vec{P}_{SGT} = \frac{e}{Mc}\left\{\begin{array}{l}\left(\frac{g}{2}-\frac{\gamma}{\gamma+1}\right)\left(\vec{s}\times\vec{E}/c\right) - \frac{\gamma^3\vec{\beta}}{(\gamma+1)^2}\left[\vec{\beta}\left(\vec{s}\times\vec{E}/c\right)\right] - \gamma\vec{\beta}\left(\vec{s}\vec{B}\right) - \\ -\left(\frac{g}{2}-1\right)\left[\frac{\gamma^3\vec{\beta}}{(\gamma+1)^2}\left(\vec{\beta}\vec{s}\right)\left(\vec{\beta}\vec{B}\right) + \frac{\gamma}{\gamma+1}\left(\vec{s}\left(\vec{\beta}\vec{B}\right) + \vec{B}\left(\vec{\beta}\vec{s}\right)\right)\right]\end{array}\right\} \quad . \tag{4}$$

## 4. Stern-Gerlach-Thomas Force

The Lagrange equation of motion states that

$$d\vec{P}/dt = \vec{\nabla}L = \left(\partial L/\partial x_1; \partial L/\partial x_2; \partial L/\partial x_3\right) \quad . \qquad (5)$$

From eqs. (3) and (5), we obtain

$$Mc\ d(\gamma\vec{\beta})/dt \ = \vec{F}_L + \vec{F}_{SGT} \qquad (6)$$

where $\vec{F}_L$ is the Lorentz force

$$\vec{F}_L = e\vec{E} + ec\vec{\beta}\times\vec{B} \quad ,$$

and $\vec{F}_{SGT}$ the Stern-Gerlach-Thomas force

$$\vec{F}_{SGT} \equiv \vec{\nabla}L_{SGT} - d\vec{P}_{SGT}/dt \ . \qquad (7)$$

Before we evaluate eq.(7) from eqs. (2) and (4), it is useful to comment on a practical application of the SGT force. Because this force depends linearly on the particle spin $\vec{s}$ , it was proposed [1] to use this force to polarize an un-polarized particle beam or alternatively to measure the polarization of a beam (polarimeter). In either case, the force is applied or measured by passing a stored particle beam through a set of radio-frequency cavities.

However if a particle traverses a localized field region (zero field outside the region), it can be seen from eqs.(6) and (7) that the change in mechanical momentum $Mc\gamma\vec{\beta}$ , i.e. the integral

$$\Delta\left(Mc\gamma\vec{\beta}\right) = \int_{field\ region} dt\cdot Mc\cdot d\left(\gamma\vec{\beta}\right)/dt$$

is not affected by the SGT force term $d\vec{P}_{SGT}/dt$ because it is a total differential in time and $\vec{P}_{SGT}$ is zero outside the field region. In this case, the only contributing SGT force is $\vec{\nabla}L_{SGT}$ where

$$\frac{\partial L_{SGT}}{\partial x_k} = \frac{e}{M}\left\{\left(G2+\frac{1}{\gamma+1}\right)\vec{\beta}\left(\vec{s}\times\frac{\partial\vec{E}/c}{\partial x_k}\right) - G2\frac{\gamma}{\gamma+1}\left(\vec{\beta}\vec{s}\right)\left(\vec{\beta}\frac{\partial\vec{B}}{\partial x_k}\right) + \left(G2+\frac{1}{\gamma}\right)\left(\vec{s}\frac{\partial\vec{B}}{\partial x_k}\right)\right\}$$

and

$$G2 \equiv \frac{g}{2} - 1 \quad .$$

For completeness, we calculate the total SGT force of eq. (7). Since $\vec{P}_{SGT}$, in analogy to the electromagnetic momentum $\vec{P}_{EM} = e\vec{A}$, is treated as a function of $\vec{x}$ and $t$ only, we have

$$\begin{aligned}\vec{F}_{SGT} &= \vec{\nabla}L_{SGT} - d\vec{P}_{SGT}/dt = -\vec{\nabla}H_{SGT} + \vec{\nabla}\left(c\vec{\beta}\cdot\vec{P}_{SGT}\right) - c\left(\vec{\beta}\cdot\vec{\nabla}\right)\vec{P}_{SGT} - \partial\vec{P}_{SGT}/\partial t = \\ &= -\vec{\nabla}H_{SGT} - \partial\vec{P}_{SGT}/\partial t + c\vec{\beta}\times\left(\vec{\nabla}\times\vec{P}_{SGT}\right)\end{aligned}$$

where $H_{SGT} = c\vec{\beta}\cdot\vec{P}_{SGT} - L_{SGT}$ is the SGT-component of the Hamiltonian.

Since, according to eq. (3)

$$c\frac{\partial P_{SGT,k}}{\partial t}=\frac{\partial}{\partial t}\left(\frac{\partial L_{SGT}}{\partial\beta_k}\right)=\frac{\partial}{\partial t}\left(\frac{\left(\vec{s}\cdot\vec{\Omega}\right)}{\partial\beta_k}\right)=\frac{\partial}{\partial t}\left(\frac{\partial\vec{s}}{\partial\beta_k}\cdot\vec{\Omega}+\vec{s}\cdot\frac{\partial\vec{\Omega}}{\partial\beta_k}\right)=$$

$$=\frac{\partial}{\partial\beta_k}\left(\frac{\partial\vec{s}}{\partial t}\right)\cdot\vec{\Omega}+\frac{\partial\vec{s}}{\partial\beta_k}\cdot\frac{\partial\vec{\Omega}}{\partial t}+\frac{\partial\vec{s}}{\partial t}\cdot\frac{\partial\vec{\Omega}}{\partial\beta_k}+\vec{s}\cdot\frac{\partial^2\vec{\Omega}}{\partial\beta_k\partial t}$$

where, from eq. (1), $\partial\vec{s}/\partial t=\vec{s}\times\vec{\Omega}$, and $\partial\vec{s}/\partial\beta_k=0$, we find

$$c\frac{\partial P_{SGT,k}}{\partial t}=\left(\vec{s}\times\frac{\partial\vec{\Omega}}{\partial\beta_k}\right)\cdot\vec{\Omega}+\left(\vec{s}\times\vec{\Omega}\right)\cdot\frac{\partial\vec{\Omega}}{\partial\beta_k}+\vec{s}\cdot\frac{\partial^2\vec{\Omega}}{\partial\beta_k\partial t}=\vec{s}\cdot\frac{\partial^2\vec{\Omega}}{\partial\beta_k\partial t}$$

i.e. the precession $\partial\vec{s}/\partial t$ of the spin does not contribute to the time variation $\partial\vec{P}_{SGT}/\partial t$ of $\vec{P}_{SGT}$.

Assuming, for simplicity, $\vec{\beta}$ to point in the $x_3$-direction, such that, from eqs. (2) and (4),

$$\frac{Mc}{e}P_{SGT1,2}=\left(\frac{g}{2}-\frac{\gamma}{\gamma+1}\right)\Sigma_{1,2}-G2\frac{\gamma\beta}{\gamma+1}\left(s_{1,2}B_3+s_3B_{1,2}\right);$$

$$\frac{Mc}{e}P_{SGT3}=\left(\frac{g}{2}-\frac{\gamma^2}{\gamma+1}\right)\Sigma_3-\gamma\beta\left(\vec{s}\vec{B}\right)-G2\cdot\gamma\beta s_3B_3;$$

$$\frac{M}{e}H_{SGT}=-\frac{\gamma\beta\left(\gamma-1\right)}{\gamma+1}\Sigma_3-\left(G2+\gamma\right)\left(\vec{s}\vec{B}\right)-G2\left(\gamma-1\right)s_3B_3$$

where we have defined $\vec{\Sigma}\equiv\vec{s}\times\vec{E}/c$, we find

$$\frac{M}{e}F_{SGT1,2}=-\frac{M}{e}\partial H_{SGT}/\partial x_{1,2}-\frac{M}{e}\partial P_{SGT1,2}/\partial t\mp\frac{Mc}{e}\beta\left(\vec{\nabla}\times\vec{P}_{SGT}\right)_{2,1}=$$

$$=\left(\frac{g}{2}-\frac{\gamma}{\gamma+1}\right)\left[\beta\frac{\partial\Sigma_3}{\partial x_{1,2}}-\beta\frac{\partial\Sigma_{1,2}}{\partial x_3}-\dot{\Sigma}_{1,2}\right]+\left(G2+\frac{1}{\gamma}\right)\frac{\partial\left(\vec{s}\vec{B}\right)}{\partial x_{1,2}}+$$

$$+G2\frac{\gamma\beta}{\gamma+1}\left(\beta s_{1,2}\frac{\partial B_3}{\partial x_3}+\beta s_3\frac{\partial B_{1,2}}{\partial x_3}-\beta s_3\frac{\partial B_3}{\partial x_{1,2}}+s_{1,2}\dot{B}_3+s_3\dot{B}_{1,2}\right)$$

and

$$\frac{M}{e}F_{SGT,3}=-\frac{M}{e}\partial H_{SGT}/\partial t-\frac{M}{e}\partial P_{SGT3}/\partial t=$$

$$=+\frac{\gamma\beta\left(\gamma-1\right)}{\gamma+1}\frac{\partial\Sigma_3}{\partial x_3}-\left(\frac{g}{2}-\frac{\gamma^2}{\gamma+1}\right)\dot{\Sigma}_3+\left(G2+\gamma\right)\frac{\partial\left(\vec{s}\vec{B}\right)}{\partial x_3}+\gamma\beta\left(\vec{s}\dot{\vec{B}}\right)+$$

$$+G2\left[\left(\gamma-1\right)s_3\frac{\partial B_3}{\partial x_3}+\gamma\beta s_3\dot{B}_3\right]$$

where the dots indicate 1/c times the partial derivative with respect to time and $\dot{\vec{\Sigma}}=\vec{s}\times\dot{\vec{E}}/c$ .

For highly relativistic particles, the SGT force components reduce to

$$\frac{M}{e}F_{SGT1,2}\xrightarrow[\gamma\gg 1]{}G2\left\{\frac{\partial\left(\Sigma_3+\vec{s}\cdot\vec{B}-s_3B_3\right)}{\partial x_{1,2}}+\frac{d\left(s_{1,2}B_3+s_3B_{1,2}-\Sigma_{1,2}\right)}{dt}\right\}=$$

$$=\frac{M}{e}G2\left[\partial L_{SGT}/\partial x_{1,2}-dP_{SGT1,2}/dt\right]$$

and

$$\frac{M}{e}F_{SGT,3}\xrightarrow[\gamma\gg 1]{}\frac{d}{dt}\left[\gamma\left(\Sigma_3+\vec{s}\cdot\vec{B}+G2\cdot s_3B_3\right)-\left(G2+2\right)\Sigma_3\right]+G2\cdot\partial\left(\Sigma_3+\vec{s}\cdot\vec{B}-s_3B_3\right)/\partial x_3=$$

$$=\frac{M}{e}\left[G2\cdot\partial L_{SGT}/\partial x_3-dP_{SGT3}/dt\right]$$

plus terms of order ($1/\gamma$), where

$$d/dt\xrightarrow[\gamma\gg 1]{}\partial/\partial t+\partial/\partial x_3 \quad +\text{ terms of order }\left(1/\gamma^2\right) .$$

The leading term of the $x_3$-component, proportional to $\gamma$, is a total time derivative and the SGT force does not contain any $\gamma^2$-terms.

In the rest frame ( $\vec{\beta}\to 0$ ), the SGT force becomes

$$\frac{M}{e}\vec{F}_{SGT}\xrightarrow[\beta\to 0]{}\frac{g}{2}\vec{\nabla}\left(\vec{s}\vec{B}\right)-\frac{g-1}{2}\left(\vec{s}\times\dot{\vec{E}}/c\right)$$

and

$$\frac{Mc}{e}\vec{P}_{SGT}\xrightarrow[\beta\to 0]{}\left(\frac{g-1}{2}\right)\left(\vec{s}\times\vec{E}/c\right); \qquad \frac{M}{e}L_{SGT}\xrightarrow[\beta\to 0]{}\left(\frac{g-1}{2}\right)\vec{\beta}\cdot\left(\vec{s}\times\vec{E}/c\right)+\frac{g}{2}\left(\vec{s}\vec{B}\right)$$